\documentclass[11pt]{article}
\usepackage[margin=1in]{geometry}
\usepackage{mathptmx}
\usepackage[T1]{fontenc}
\usepackage[utf8]{inputenc}
\usepackage{graphicx}
\usepackage{float}
\usepackage{amsmath}
\usepackage{setspace}
\usepackage[numbers,sort&compress]{natbib}
\usepackage{microtype}
\usepackage[hidelinks]{hyperref}
\usepackage{caption}
\usepackage{booktabs}
\graphicspath{{figures/}}

\title{\textbf{Who Anchors AI Overviews in Health? Baidu, Google, and the Geography of Authority}\thanks{Correspondence: \href{mailto:mingyue.zha.27@dartmouth.edu}{mingyue.zha.27@dartmouth.edu}.}}
\author{Mingyue Zha\\ \small Program in Quantitative Social Science, Dartmouth College, Hanover, NH 03755
\and Ho-Chun Herbert Chang\\ \small Program in Quantitative Social Science, Dartmouth College, Hanover, NH 03755}
\date{}

\begin{document}
\maketitle

\begin{abstract}
\noindent Artificial intelligence is being rapidly incorporated into traditional search systems, yet scant work audits the information disparities across platforms, geography, and languages. We address this gap by comparing Google and Baidu's AI Overview systems for health queries, and measure informational anchors that emerge. Auditing 1,920 health queries across 12 countries and 4 languages, we find that Google and Baidu exhibit vertical integration, routing users toward their own company platforms in AI Overviews rather than a diverse set of primary sources. Smaller, lower-localization countries receive fewer domestically sourced references for health queries. Issuing the same query in a country's official language rather than English raises the share of locally sourced citations approximately 3.5- to 13.5-fold. Comparing queries across health topics of varying severity and controversy, including Traditional Chinese Medicine as an example, we also show that health disclaimers are multidimensional and vary across language and culture. We discuss how generative search influences access to health information, and the urgent need for culturally-aware oversight of these systems that influence critical health decisions.

\end{abstract}

\section{Introduction}

Every day, Google processes approximately sixteen billion search queries, over one billion of which concern healthcare \cite{statcounter2025worldwide}. Search is arguably one of the most consequential applications of generative artificial intelligence (AI) \cite{aral2026rise}. People rely on web search for information to make critical decisions, including decisions that amount to matters of life or death. The information these systems present and the way they are framed can influence medical decisions, treatment adherence, and perceptions of contested therapies \cite{moorhead2013social,holone2016filter}.

With the advent of large language models, search engines integrated AI into their platforms, with Google and Bing AI Overviews launching in May and July 2024 respectively \cite{reid2024generative,bing2024generative}. From 2024 to 2025, overall exposure to Google AI Overviews (AIO) expanded from 7 to 229 countries, due to Google's dominance as a global search engine \cite{aral2026rise}. However, one country remains excluded from Google's dynamics: China. China's Great Firewall regulates and censors domestic internet, including Google and all of its related products \cite{comparitech2026google}. Instead, Chinese netizens use Baidu, a major Chinese company often called the ``Google of China'', which offers a range of online services, mobile apps, and AI technologies. In terms of market capture, Baidu commands 63.97\% of the search market for 1.1 billion Chinese internet users \cite{lee2026popular}. However, Baidu accounts for less than 1\% of the total search market outside of China \cite{wright_baidu_2024}. This has made Baidu region-specific and localized for Chinese users.

\textbf{AI Overviews and information environments.} Unlike a ranked list of links, an AI Overview delivers a single pre-synthesized answer positioned above every organic result. It promises to lower the cognitive burden of a search but narrows a user's exposure to source diversity \cite{sharma2024generative,stadler2024cognitive}. Because of this placement, AI Overviews carry an out-sized influence over public knowledge, opinions, and people's behavioral intentions \cite{aral2026rise,xu2025aisummaries}.

Google's AI Overviews reduce outbound organic clicks by roughly 39.8\% and raise the rate of zero-click searches by 34.5\% when the feature appears \cite{agarwal2026googleaio}, with actual visits to an AI Overview source in only about 1\% of cases \cite{chapekis2026clickbehaviors}. Critically, biases carried by the model from its training data or its retrieval process may propagate to billions of users \cite{urman2022matter}. Indeed, the reliability of AI in search has been brought into question. Large language models can produce fluent, confident, and factually unsupported content (i.e. hallucinations) \cite{ji2023survey}. For instance, Google's generative search engine advised users to ``eat rocks'' for nutrients and to glue cheese onto pizza in 2024 \cite{heikkila2024googleai}. After OpenAI launched ChatGPT Search, quotes were misattributed to their original sources in 134 of 200 attempts, roughly 67\% of the time \cite{jazwinska2025citing}.

\textbf{Geography, platforms, and language.} Geolocation is one of the features used by search platform algorithms, and their effects on conventional search are well documented. An audit study that varied a user's GPS coordinates across the United States found differences in search growth with physical distance \cite{klimansilver2015location}. Across studies, IP address and login state are consistently the strongest predictors of how much a result set is personalized \cite{hannak2013measuring,klimansilver2015location}. Personalization in search may be helpful for filtering for relevancy, but may also lead to informational disparities, such as uneven geographic distribution for abortion clinic results in rural areas \cite{mejova2022googling}. Fewer studies investigate this at a national level. Running 24,000 queries across 243 countries, Aral, Li and Zuo \cite{aral2026rise} found that deployment is highly unequal and not just a function of internet penetration. Audits of Google, Bing, DuckDuckGo, Yahoo, Baidu, and Yandex found substantial cross-engine differences \cite{urman2022matter}, with the overlap between Google and Bing always under 32\% for a given query \cite{fischer2022comparison}.

More importantly, the majority of existing web search literature is about Google and countries that use Google. China, which accounts for 17\% of the world's population, has a distinct information ecosystem and arguably the largest information bubble in existence. China's Great Firewall excludes most Western domains, including YouTube and Wikipedia. However, studies comparing Google and Baidu are sparse. Jiang \cite{jiang2014concentration} compared 6,320 query results across the two engines and found only 6.8\% overlap and little ranking similarity. Chinese search engines also rarely direct users beyond national borders and disproportionately favor the country's own content, with 32\% of Baidu's results linking to Baidu's own properties compared with 8\% for Google \cite{jiang2014business}.

To date, information on Baidu's AI Overview is limited largely to the platform's own account. Baidu's own search team has since published a technical description of a four-agent architecture underlying the system \cite{li2025aisearch}, but this has not been independently audited, and no published study has yet compared Baidu's generative search output with Google's on matched queries.

Language can also shape search results independently of geography or platform. For LLMs, the use of different languages can surface different levels of bias and accuracy \cite{chang2025language}. A cross-lingual study of ChatGPT, Claude, and Gemini found that political questions are moderated by language, when associated with a country that exerts heavy control over its media, where models produce answers favorable to that government \cite{yang2026crosslingual}. On Google, 53\% of English-language suicide queries surfaced crisis hotline information, compared to just 13\% of Spanish-language queries \cite{borge2021suicide}. However, there is still a dearth of comparisons in Chinese.

\textbf{Health information in AI Overviews.} As of early 2026, Google's AI Overview appears for 64.7\% of question-form queries overall \cite{xu2026measuring}. Despite this rapid rollout, few studies have examined differences and potential biases in the content of AI Overviews across countries, platforms, and languages. In this study, we investigate how AI Overviews diverge across health-related queries.

We specifically study the case of health queries for three reasons. First, broadly, search is central to health decisions, with high-stakes consequences. For instance, what parents find in response to a vaccination query is a deciding factor in whether their child becomes vaccinated or not \cite{holone2016filter}.

Second, health is a domain where scientific information is not expected to vary across countries, but may differ due to national regulations and cultural differences. Health policies and regulations actively impact public health priorities and attitudes toward specific ailments \cite{burris2002disease,henderson2013mental}. For example, following the COVID-19 outbreak, Chinese media platforms filtered coronavirus-related terms due to possible surfacing of criticism toward the government's outbreak response \cite{ruan2020censored}.

Third, different countries and cultures may have different perspectives on health and associated treatment frameworks. Traditional Chinese Medicine (``TCM'') is a medical tradition that encompasses practices including herbal medicine, acupuncture, and other physical and psychological approaches. Several of these practices have grown increasingly popular in the United States, with acupuncture use, for example, more than doubling between 2002 and 2022 \cite{nahin2024use, nccih2024rise}. TCM and Western biomedicine represent two distinct frameworks of health that we also investigate.

This study is guided by three primary research questions:
\begin{enumerate}
\item[1.] \textbf{Platform}: How do AI Overview responses to the same health queries differ between Google and Baidu?
\item[2.] \textbf{Anchoring}: How do Google AI Overview deployment rates and citations vary across English-speaking countries?
\item[3.] \textbf{Language}: How does language impact health query responses?
\end{enumerate}

\section{Results}

\subsection{Privatization of information associated with AI Overviews by platform}

Using SerpApi's search APIs, we collected 1,920 responses to health queries across 12 countries, 4 languages, and two platforms, Google and Baidu. To confirm that our results were not driven by stochastic variation, we repeated the same query multiple times per country over the course of May 2026. Responses did not vary meaningfully, and this repetition did not affect our results (see Methods for details).

\begin{figure}[H]
\centering
\includegraphics[width=.63\linewidth]{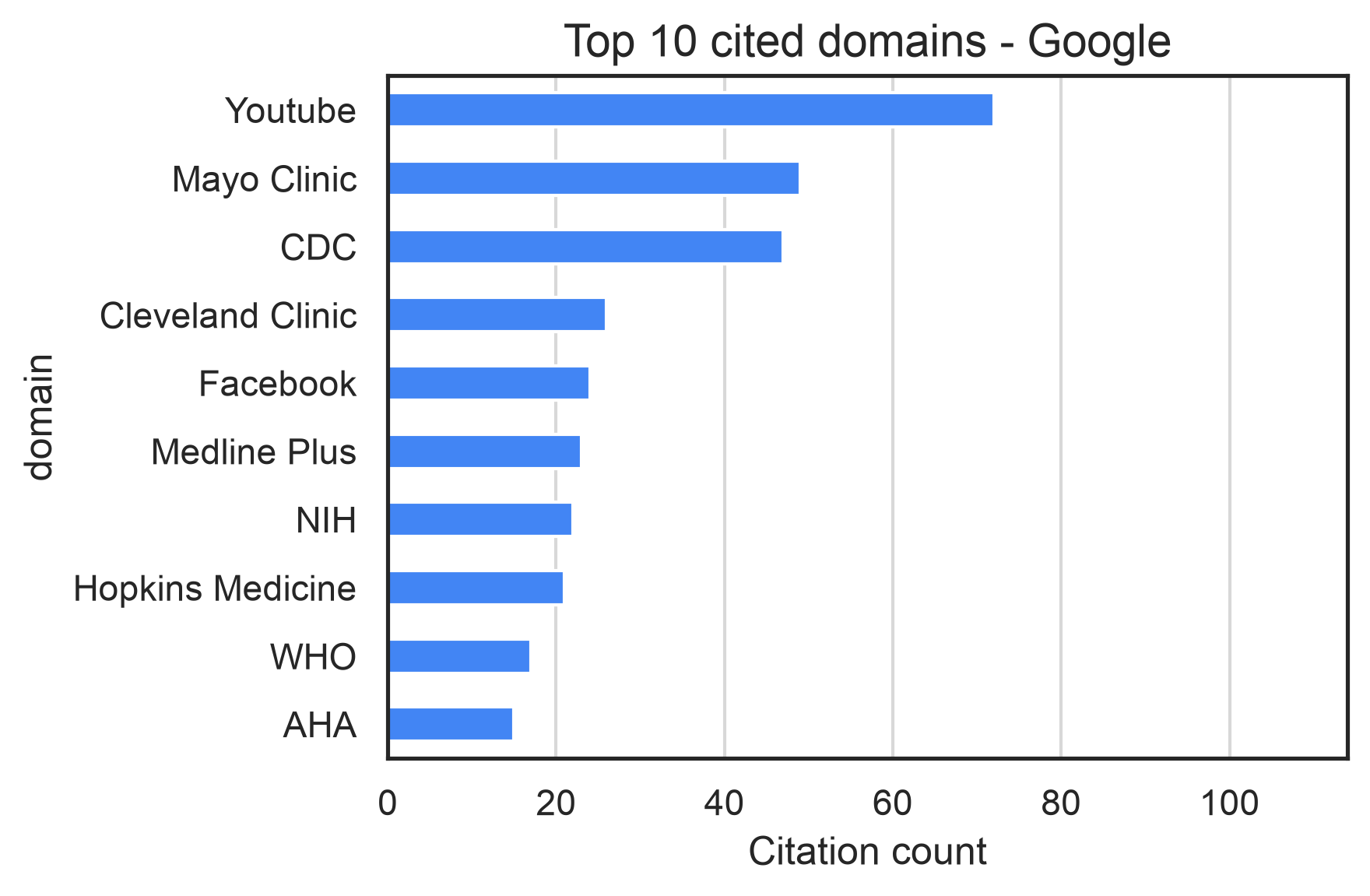}\\[4pt]
\includegraphics[width=.63\linewidth]{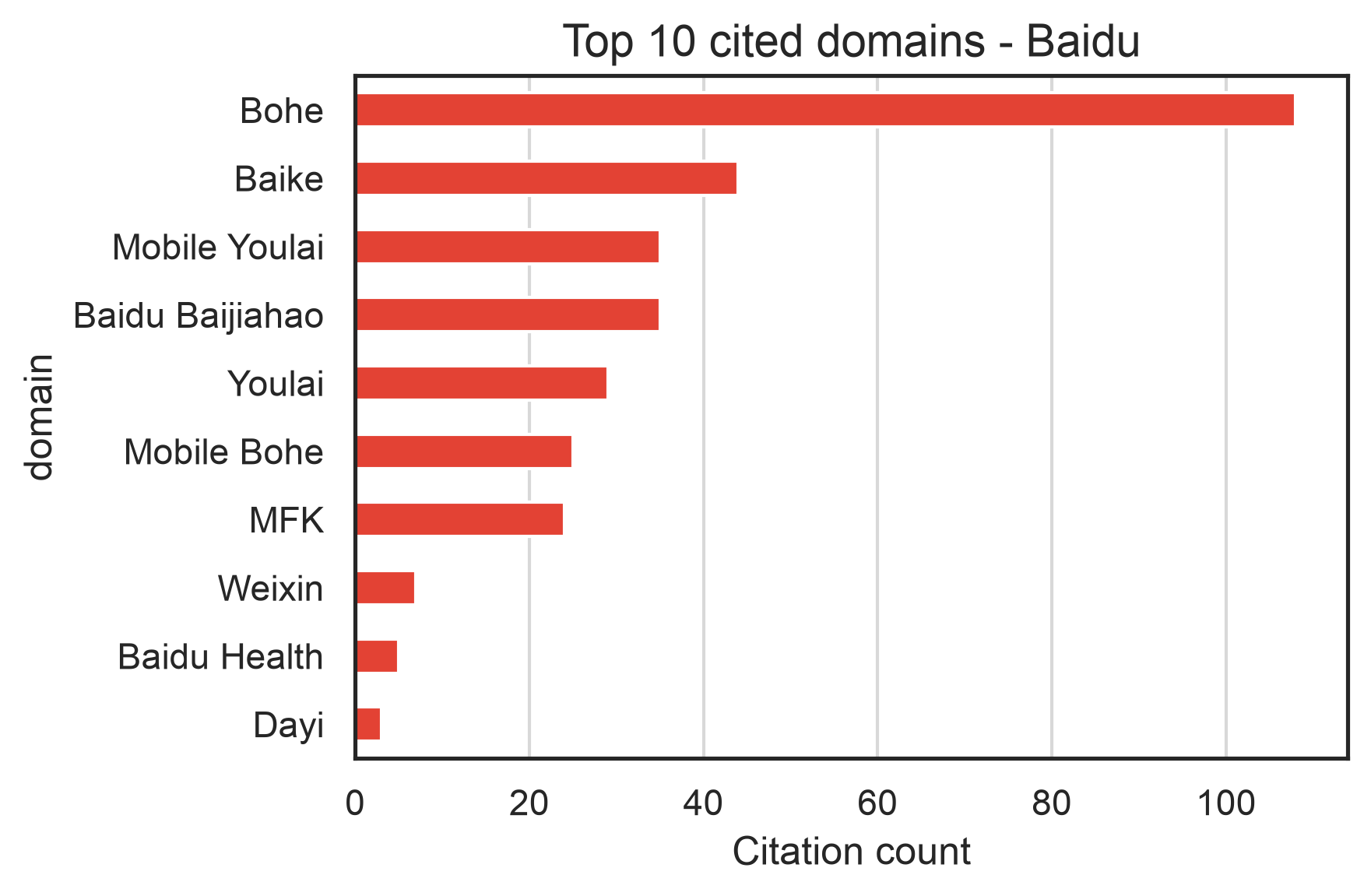}
\caption{\textbf{Top cited domains on a) Google (US) and b) Baidu (China).} YouTube is the top cited domain on Google.}
\label{fig1}
\end{figure}

First, we examined the ten most-cited domains for Google's and Baidu's health AI Overviews. Citations indicate the authorities and sources that AI Overview systems draw on to synthesize their answers, at times quoting directly from the referenced material \cite{google2025aioverviews,xu2026measuring}. They thus serve as a proxy for the types of institutions and information channels that AI Overviews rely on for health guidance. Both systems concentrated their citations on a relatively small set of intermediary platforms rather than on a diverse range of primary sources (Figure~\ref{fig1}).

Figure~\ref{fig1}a shows Google's AI Overview citations in the US. YouTube was the most frequently cited domain, exceeding established health authorities such as Mayo Clinic, the CDC, and the WHO. YouTube was also the top-cited source in many other countries. Because YouTube is owned by Google's parent company, Alphabet, directing users to YouTube may also increase engagement within Google's own ecosystem and create opportunities for advertising revenue.

Figure~\ref{fig1}b shows that on Baidu, the most frequently cited domain was Bohe, followed by Baike (Baidu's encyclopedia service), and other Baidu-operated platforms including Baidu Baijiahao and Baidu Health. Similar to Google, Baidu concentrated citations within its own corporate ecosystem. However, Bohe is unique. Bohe.cn (Chinese: Boh\'e Y\={\i}sh\=eng) is an independent Chinese online medical-information platform. It offers multiple services, including hosting articles on disease and medication information, hospital appointment directories, and health news, as well as short-form content.

Whereas Google's most-cited domain was YouTube, an open user-generated video platform, Baidu primarily referenced its encyclopedia and health information services, along with independent publishing platforms. Both systems exhibit vertical integration in AI Overview references through different intermediaries. This is consistent with prior findings that search engines favor properties tied to their own commercial interests \cite{jiang2014business}. However, the effect seems more pronounced in AI Overviews than in conventional search results. The first organic link on a typical health search is not, after all, a YouTube video.

Because YouTube functions as Google's main source for health content, we examined the top YouTube channels cited within AI Overviews across all countries. Table~\ref{tab1} shows that citations were distributed across a mix of health-system and hospital channels, led by Mass General Brigham, Temple Health, and UC Davis Health. Mayo Clinic, despite being the most-cited institutional domain in US Google search, is not in the top 10 of YouTube channels cited, while non-institutional channels such as the Infographics Show and Doctor O'Donovan are. This suggests that the authority of the cited source may depend on channels surfaced by YouTube's ranking mechanisms rather than on health-specific editorial review. As a result, reliance on YouTube may introduce a vulnerability in the credibility of AI health overviews, since the quality and authority of cited content can vary substantially even when all of them appear under the YouTube domain.

\begin{table}[H]
\centering
\caption{Top 10 cited YouTube channels in Google AI Overviews for English health queries across countries.}
\label{tab1}
\begin{tabular}{lc}
\toprule
Channel & Citation events \\
\midrule
Mass General Brigham   & 53 \\
Temple Health           & 48 \\
UC Davis Health         & 37 \\
UHNToronto              & 34 \\
Cleveland Clinic        & 32 \\
MedSurge India          & 28 \\
Horizon Health Network  & 25 \\
GoodRx                  & 24 \\
Doctor O'Donovan        & 23 \\
The Infographics Show   & 20 \\
Nucleus Medical Media   & 19 \\
\bottomrule
\end{tabular}
\end{table}

\subsection{U.S. and international institutions are cited as authorities beyond their own countries}

\begin{figure}[H]
\centering
\includegraphics[width=.85\linewidth]{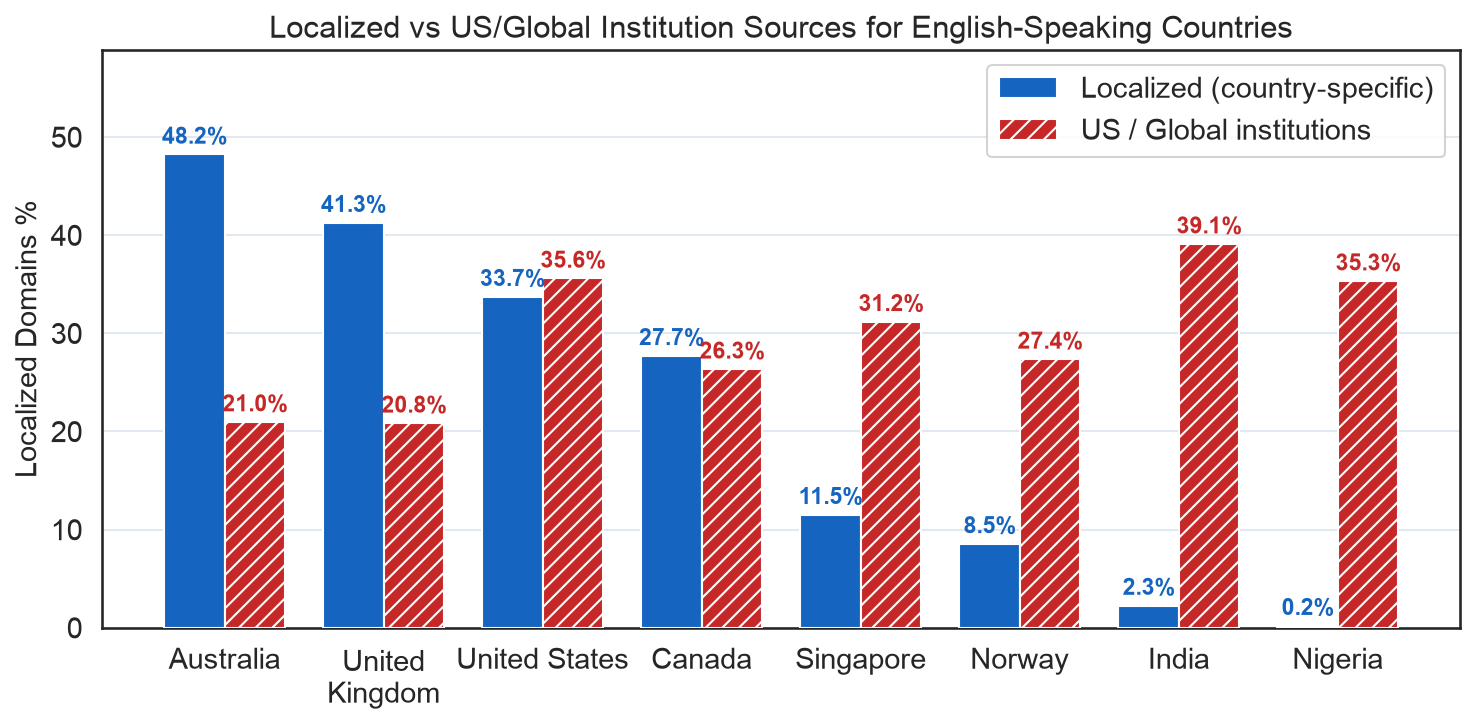}
\includegraphics[width=.53\linewidth]{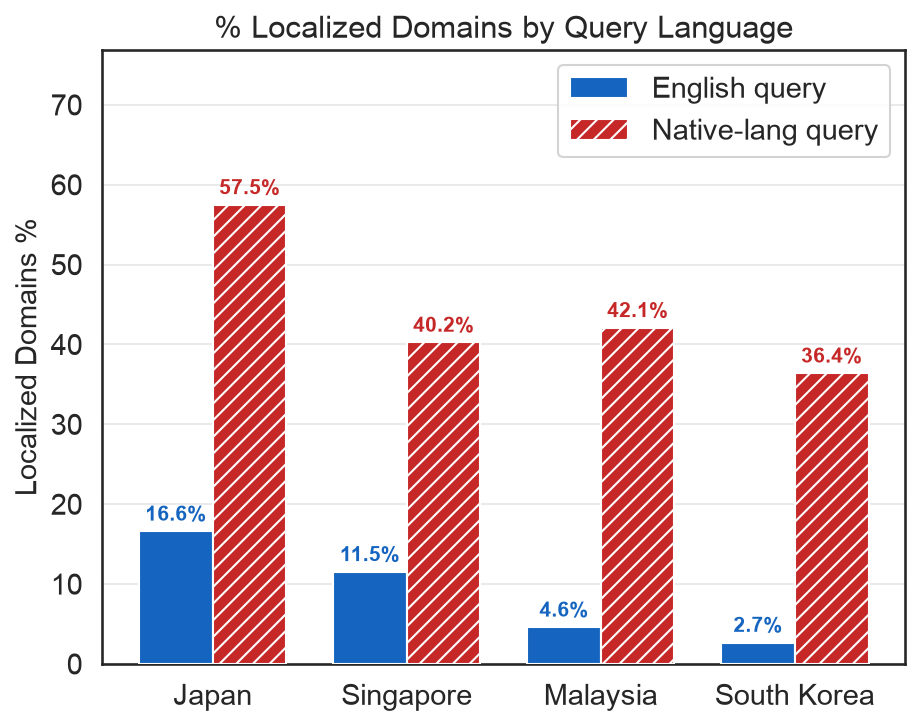}

\caption{\textbf{Percentage of institutions that are local versus US/global across countries and languages.} a) Localized (country-specific) versus U.S./global institution sources for eight countries, using English-language queries. b) Percentage of localized domains by query language (English versus native language) for Japan, Singapore, Malaysia, and South Korea.}
\label{fig2}
\end{figure}

Figure~\ref{fig2}a shows the share of cited sources that were localized to the querying country, using country-specific top-level domains, against the share sourced from the United States and other international institutions. Localization was highest for Australia (48.2\%) and the United Kingdom (41.3\%), intermediate for the United States (33.7\%) and Canada (27.7\%), and lowest for Singapore (11.5\%), Norway (8.5\%), India (2.3\%), and Nigeria (0.2\%). Overall, countries outside this set of four large, high-localization markets received a smaller proportion of domestically sourced citations. Singapore, Norway, India, and Nigeria were much more likely to receive responses citing U.S. or other international institutions than sources from their own countries. These findings suggest that AI Overviews frequently prioritize established U.S. and international health institutions over local sources in smaller markets, even when responding to queries submitted from countries with their own established national health authorities.

To separate the effects of query language and geography, we compared the proportion of localized sources returned for Japan, Singapore, Malaysia, and South Korea when identical health queries were issued in English versus each country's primary native language. Figure~\ref{fig2}b shows that localized sourcing was consistently low when queries were issued in English, accounting for 16.6\% of citations in Japan, 11.5\% in Singapore, 4.6\% in Malaysia, and 2.7\% in South Korea. In contrast, issuing the same queries in the country's native language substantially increased the share of local sources, reaching 57.5\% in Japan, 42.1\% in Malaysia, 40.2\% in Singapore, and 36.4\% in South Korea. Across countries, this represents an approximately 3.5- to 13.5-fold increase in localization, ranging from a 3.5-fold increase in Japan and Singapore to a 9-fold increase in Malaysia and a 13.5-fold increase in South Korea. This indicates that query language is an important determinant of whether locally relevant sources show up in AI Overview citations.

Consequently, users who search for health information in English, including local users who do so out of habit, educational background, or professional necessity, are less likely to receive information from the local country's health information ecosystem, regardless of where the query is submitted geographically.

\subsection{Country and topics impact localization of citations}

\begin{figure}[H]
\centering
\includegraphics[width=.65\linewidth]{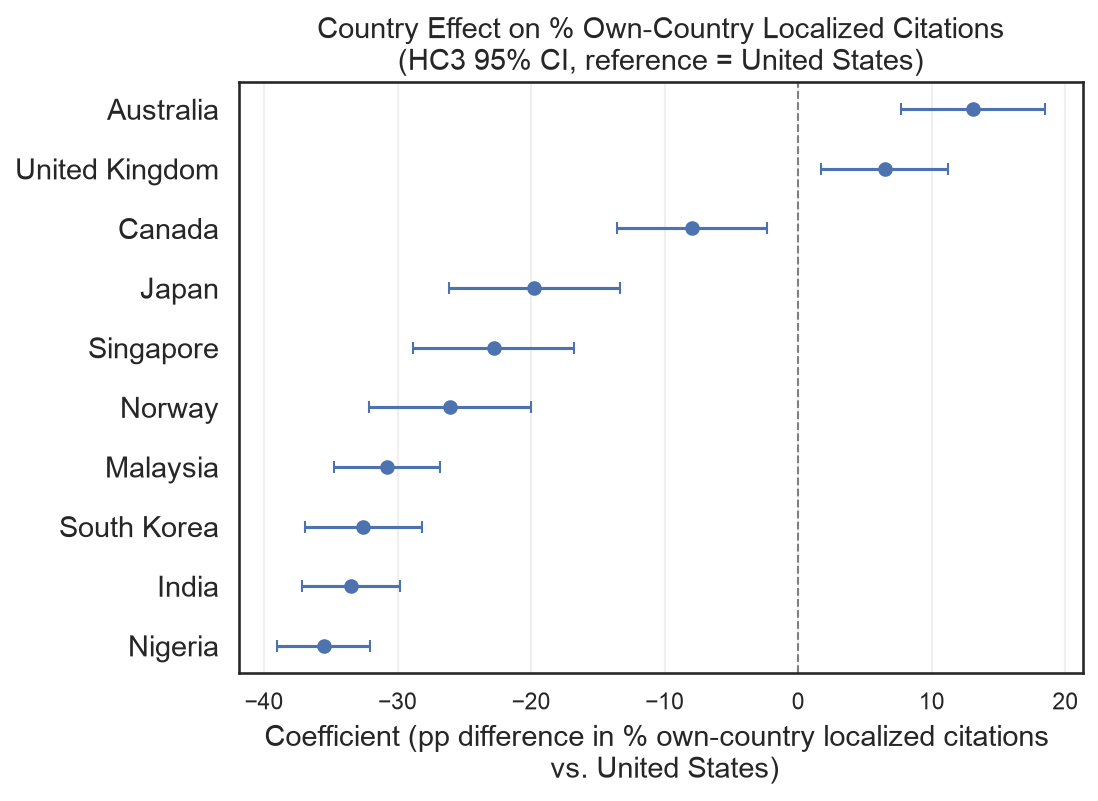}\\[4pt]
\includegraphics[width=.65\linewidth]{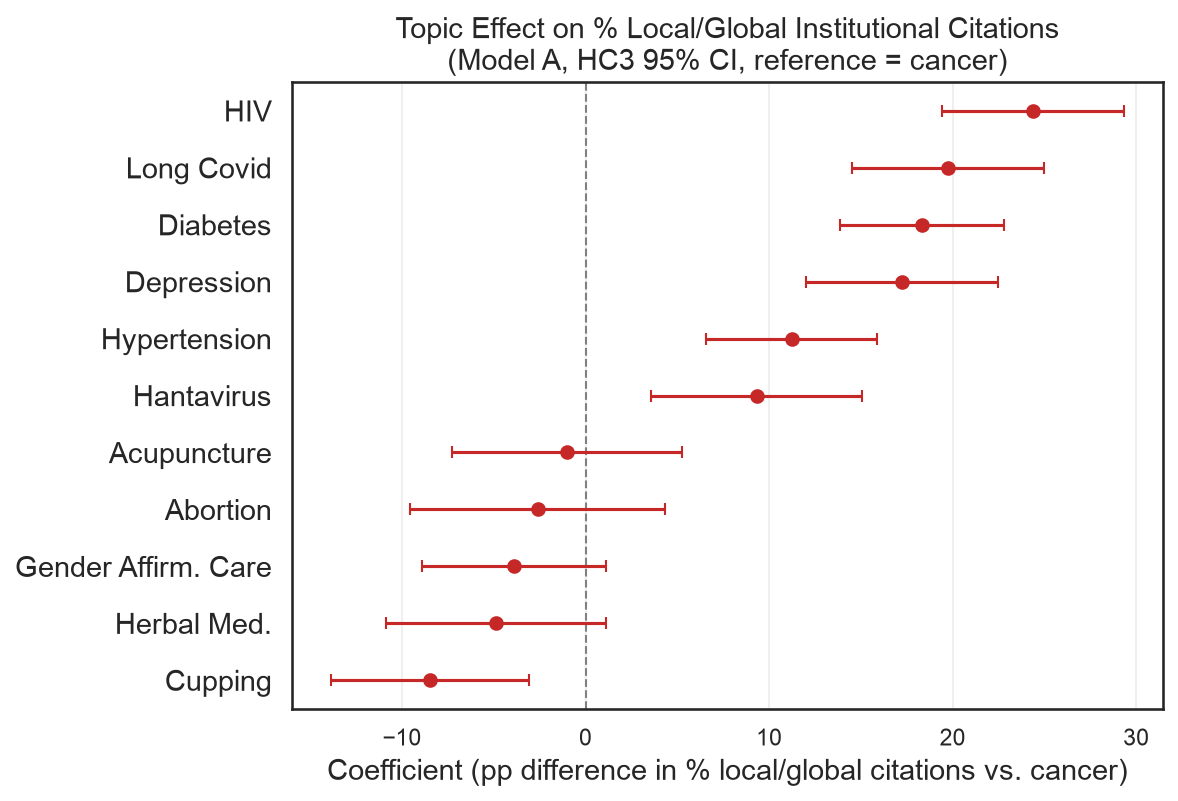}
\caption{\textbf{Country and topic effects on \% of localized sources.} a) Country effect on percentage own-country localized citations (HC3 95\% CI, reference = United States). Coefficients represent the percentage-point difference in own-country localized citations relative to the United States. b) Topic effect on percentage local/global institutional citations (Model A, HC3 95\% CI, reference = cancer). Coefficients represent the percentage-point difference in local/global institutional citations relative to cancer.}
\label{fig3}
\end{figure}

Using OLS regressions, we modeled the share of local versus global institutional citations as a function of querying country and health topic, using heteroskedasticity-robust standard errors (HC3) with the United States and cancer, respectively, as reference categories (Fig.~\ref{fig3}). The model fitting both country and topic to predict localization achieved $R^2 = 0.492$.

Figure~\ref{fig3}a plots the estimated country coefficients, relative to the United States, across the ten other countries with English query data in our sample (Australia, the United Kingdom, Canada, Singapore, Norway, India, and Nigeria, plus Japan, Malaysia, and South Korea). Australia and the United Kingdom again showed significantly positive coefficients, confirming that these countries received an even greater share of domestically sourced citations than the United States baseline. Canada's coefficient was smaller than the US. The remaining countries---Japan, Singapore, Norway, Malaysia, South Korea, India, and Nigeria---showed significantly negative coefficients, with Nigeria and India having the largest deficits. This corroborates Figure~\ref{fig2}a presented in the previous section, but is demonstrated here using a different analytical method, OLS regression.

Relative to cancer, common chronic and infectious-disease-adjacent conditions, HIV, long COVID, diabetes, depression, hypertension, and hantavirus received significantly more localized citations. Some health topics that are more contested or controversial moved the other way, though the result is only statistically significant for cupping. Certain conditions receive more localized citations than others. This may be because these conditions have more established, localized institutional resources and support. Topics related to Traditional Chinese Medicine, such as acupuncture, herbal medicine, and cupping, have comparatively less localized guidance. Interestingly, abortion and gender-affirming care also receive relatively less localized guidance, perhaps because they are politically or legally sensitive topics.

\subsection{Google and Baidu diverge in Traditional Chinese Medicine}

\begin{figure}[H]
\centering
\includegraphics[width=1.1\linewidth]{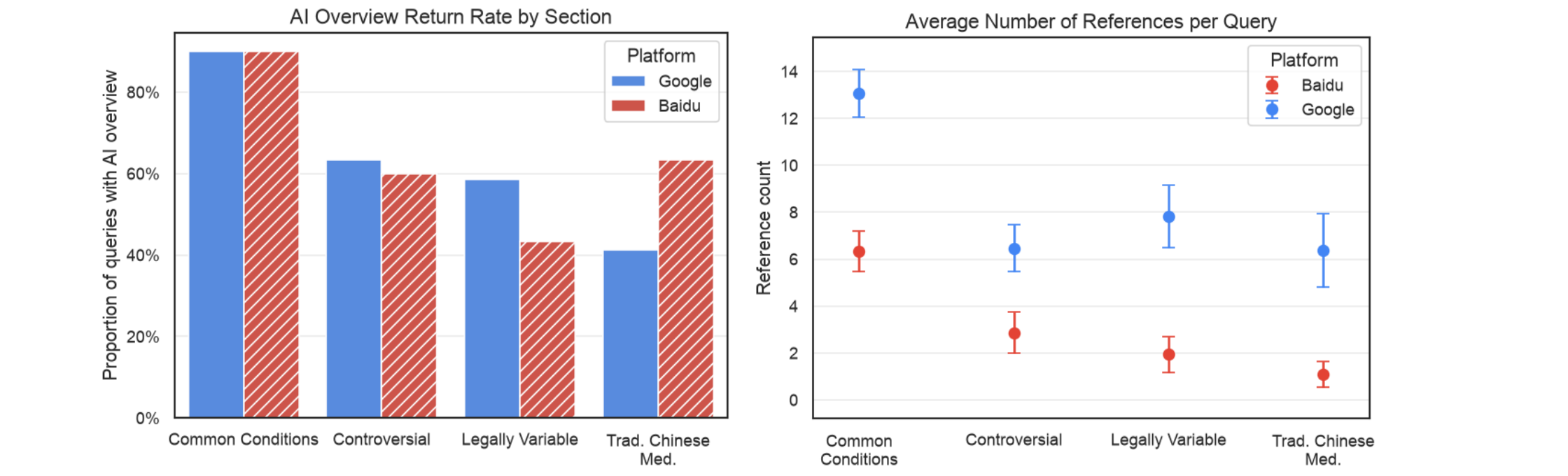}\\[4pt]
\includegraphics[width=0.85\linewidth]{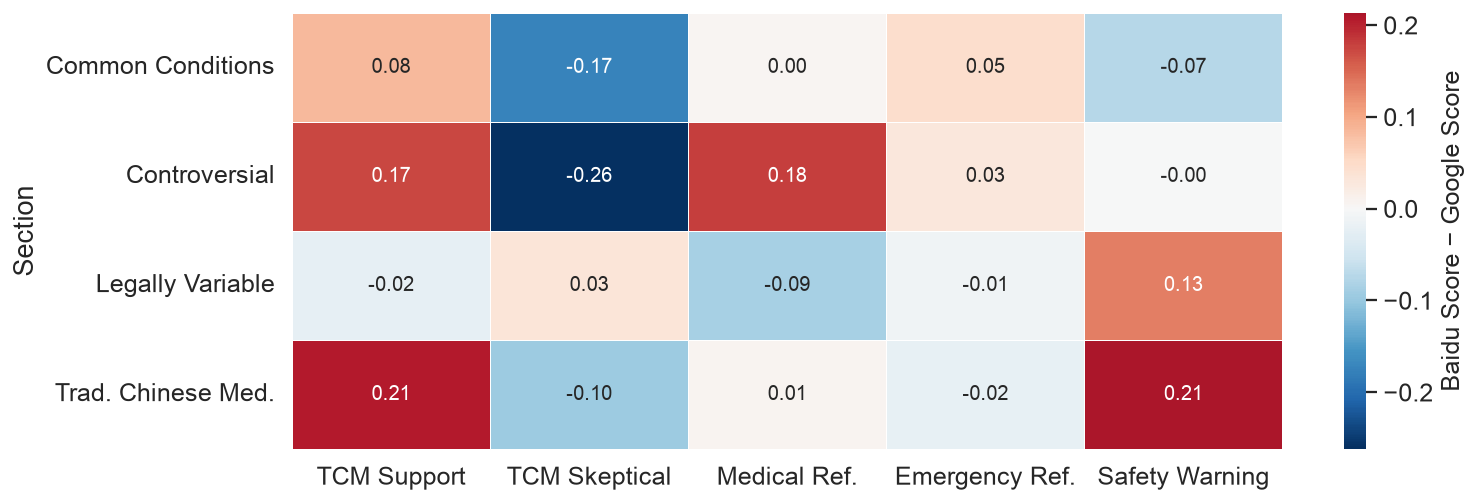}
\caption{a) AI Overview return rate by query section for Google and Baidu. b) Average number of references per query by section and platform (error bars show 95\% confidence intervals). c) Difference between Baidu and Google zero-shot label scores (Baidu score minus Google score) across query sections (rows) and features (columns): TCM support, TCM skepticism, medical referral, emergency referral, and safety warning. Red indicates Baidu scored higher on average for that feature; blue indicates Google scored higher.}
\label{fig4}
\end{figure}

We compared how consistently each platform generated an AI Overview, and how heavily each overview was sourced, across four query categories spanning health issues of varying sensitivity: common conditions, controversial topics, legally variable topics (e.g., queries whose correct answer depends on local law or regulation), and Traditional Chinese Medicine. Google's return rate declined steadily as topic sensitivity increased, from roughly 90\% of queries for common conditions down to roughly 42\% for TCM queries. Baidu's return rate followed a less consistent pattern: it matched Google's high rate for common conditions, dipped to its lowest point for legally variable topics, and then rose again for TCM queries, where it exceeded Google's return rate (Figure~\ref{fig4}a).

Averaged per health category, Google's AI Overviews drew on more source references than Baidu's overviews in every category (Figure~\ref{fig4}b). Combined with the return-rate result, this suggests that Baidu is often as likely, or more likely, than Google to generate an AI Overview for a given query, but cites substantially fewer sources when it does so, whereas Google's overviews are typically better-cited but less consistently generated for more sensitive topics.

Lastly, we used zero-shot labeling with Ollama to score responses to topic sections on a variety of features, like whether the response was supportive or skeptical of Traditional Chinese Medicine, whether it had a medical referral, emergency referral, or safety warning. Figure~\ref{fig4}c visualizes this, with red meaning Baidu on average had higher score for that feature, and blue meaning Google on average had the higher score for that feature across the section. Baidu was more extreme on TCM than Google: It was more supportive, but also featured clearer safety warnings compared to Google.

\section{Discussion}

In this study, we provide a systematic comparison of generative search across Google and Baidu for health information. We identify three crucial anchors to AI Overviews: corporate ownership, geography, and language.

First, generative search mediates access to information through platform-specific ecosystems. On Google's AI Overview, the most cited domain for health queries was YouTube, ahead of the CDC, the WHO, and the Mayo Clinic. On Baidu, the top domains included Baike and other Baidu-owned properties, but the most-cited source was an independent third-party health platform, Bohe. This suggests that the shift to generative search has been generally accompanied by the tendency of platforms to favor their own content and properties. Baidu routes users to more encyclopedic sources while Google's YouTube is an open platform where anyone can post content. This means that for YouTube, authority of the source varies greatly (from medical institutions like Mass General Brigham to general interest channels with no health credentials). Because those ecosystems differ in ownership structure and source composition, platform choice has a significant impact on where health information is synthesized from.

In testing for geographical impact, we held language constant and issued queries in English across all eight countries. Geographical localization seemed to largely favor four large, established markets: Australia, the UK, the US, and Canada had significantly higher proportions of localized domains than Singapore, Norway, India, and Nigeria. Localization also varied substantially by health topic. Conditions with well-established local health infrastructure, such as HIV, long COVID, diabetes, and depression, received a greater proportion of localized citations, whereas legally contested or politically sensitive topics---including abortion and gender-affirming care, as well as several Traditional Chinese Medicine practices---received fewer localized sources.

There may be two mechanisms causing the observed differences. First, differences in the availability, prominence, and digital visibility of authoritative local health websites constrain the extent to which search systems can localize their responses. Second, for more contentious or politically sensitive topics, or in the case of smaller-market countries, algorithms may choose to defer to more generalist or internationally recognized sources. Our findings cannot distinguish between the two explanations, and both may be contributing to the variation in localization to a certain extent.

When the country was held constant and only language changed from English to the country's official language, the share of citations drawn from that country's own health sources rose significantly. Consistent with prior literature, language plays a large role in where LLMs synthesize their information from. In practice, many people search in English out of habit, education, or professional necessity. This may inadvertently route them away from local health resources and advice even when querying in countries with robust health information infrastructures. Conversely, the increased localization when users query in their country's official language suggests that multilingual search could serve as a mechanism for improving access to locally relevant health information.

Lastly, we compared how Traditional Chinese Medicine (TCM) was handled across Baidu and Google. Baidu was generally more supportive of TCM than Google, but it also provided clearer and more explicit safety warnings. This suggests that endorsement and caution for cultural health frameworks do not necessarily oppose each other in AI Overviews. The platforms also differed in how they deployed and sourced generative responses. Google appeared less likely to provide AI Overviews for TCM queries, potentially reflecting greater caution in generating synthesized responses for health topics where evidence or consensus may be contested. Baidu, by contrast, generally generated responses but provided substantially fewer citations---an average of approximately one citation per TCM response.

This finding connects to a broader literature on how LLMs handle culturally, politically, and scientifically contested topics. Prior work suggests that model behavior can vary substantially across systems and languages when responding to sensitive topics \cite{urman2025silence, choudhary2025political, perez-toro2025exploring}. In accordance with our findings, ``health caution'' in AI Overviews can be seen as a multidimensional construct encompassing tone, safety warnings, citation quantity and quality, and willingness to generate an answer. Future evaluations of health-related generative search could similarly assess these factors as different aspects of health information quality.

\textbf{Limitations and Future Work.} First, this study captures a snapshot of two fast-moving systems, and AI Overview behavior on both platforms is likely to change as the underlying models and retrieval pipelines are updated; our findings serve as a starting point for future audit studies rather than a permanent characterization. Second, our citation analysis operates primarily at the domain level and therefore cannot fully capture the quality, accuracy, or authority of the specific content being cited. A more granular evaluation of the underlying sources would require benchmarking health claims across multiple languages and accounting for differences in national health systems, regulations, and cultural attitudes toward health and medicine. Future work can assess the factual validity, medical accuracy, evidentiary quality, and potential harms of the information generated by AI Overviews.

\section{Conclusion}

Our work offers several new insights into how generative search impacts access to health information. We show that the sources cited by AI Overviews vary by platform, country, language, and health topic. Healthcare is not a monolithic science. How healthcare is organized and practiced is often reflective of the history and values of the society delivering it. For instance, Taiwan's national health insurance formally covers Chinese medicine alongside Western medicine, a coexistence dating to 1958 \cite{wang2016effectiveness}. Physicians in China and Taiwan often combine biomedical and traditional approaches rather than treating them as mutually exclusive systems \cite{tcm2024deployed}. Medical anthropologists have long argued that even Western ``biomedicine'' itself is a culturally situated system, not a value-free default against which other traditions are measured \cite{kleinman1978culture, kleinman1997whatis}.

As AI Overviews continue to expand, we hope this study contributes to a more informed conversation about how generative search should be evaluated and governed as a health information intermediary. Decisions about who can be a health authority should be made more deliberately rather than left to the incidental effects of platform ownership, query language, and retrieval pipelines. Making sure generative responses are reflective of relevant and local health contexts is important for everyone who uses these systems, and it remains particularly important for languages and populations that have been historically under-served in global information systems.

\section{Methods}

\subsection*{Query Selection}

We selected 12 health topics spanning four categories, chosen to capture variation in how contested, stigmatized, or legally regulated a topic is (Table~\ref{tab:health_topics}). For each topic, we used Google Trends to identify the ten most searched queries in May 2026, giving $12 \times 10 = 120$ unique queries.

For the three non-English languages in our sample (Chinese, Korean, and Japanese), we translated all 120 queries into the corresponding language using Claude Sonnet 5, and had native speakers review the translations for accuracy before use.

\begin{table}[H]
\centering
\caption{Health topic categories and representative topics included in the audit.}
\label{tab:health_topics}
\small
\begin{tabular}{p{3.7cm} p{5.2cm} p{6.5cm}}
\toprule
\textbf{Category} & \textbf{Topics} & \textbf{Rationale} \\
\midrule
Common Conditions & Hypertension, Cancer, Diabetes & High global burden; established clinical guidance available \\
Controversial/Stigmatized & Depression, Long COVID, Hantavirus & Conditions with contested etiology or associated social stigma \\
Legally Variable & Abortion, Gender-Affirming Care, HIV & Legal status or clinical guidelines differ across jurisdictions \\
Traditional Chinese Medicine & Herbal Medicine, Acupuncture, Cupping & Recognition differs between China and Western nations \\
\bottomrule
\end{tabular}
\end{table}

\subsection*{Search Result Collection}

We collected search results using SerpApi's Google and Baidu Search Results API (\url{https://serpapi.com}). To simulate searches originating from each country for Google, we set the location parameter of the API request accordingly. When a Google or Baidu search returned an AI Overview, the API response included a corresponding content block, which we saved as the search result. We recorded whether an AI Overview appeared and extracted the text of that block for further analysis.

We examined 12 countries in total: Australia, Canada, China (queried via Baidu), India, Japan, Malaysia, Nigeria, Norway, Singapore, South Korea, the United Kingdom, and the United States. These countries were chosen to span a broad range of geographic regions (North America, Europe, Sub-Saharan Africa, and East, South, and Southeast Asia), market sizes, and established healthcare systems, while also including both native English-speaking and non-native-English-speaking populations, which allowed us to separate the effects of query language from those of geography. Eleven of these countries were queried in English; the twelfth, China, is excluded from Google's search ecosystem by the Great Firewall and was queried exclusively through Baidu. To further isolate the effect of language, four of the English-queried countries---Japan, Malaysia, Singapore, and South Korea---were additionally queried in their primary native language, alongside China, which was queried only in Chinese. In total, this process yielded 1,920 search results ($120 \times 11$ countries with English queries $+$ $120 \times 5$ countries with non-English queries).

To confirm that our results were not driven by stochastic variation in the search or generation process, we repeated the same query multiple times per country over the course of May 2026. Responses were often identical within a given country across repeated queries, and where they varied, the differences were minor and did not meaningfully affect our results.

\subsection*{Domain Distributions}

Across all 1,920 queries, a total of 11,944 citations were yielded. We examined the URLs of these references.

Each country in our sample has a distinct top-level URL domain, which lets us determine whether a cited source was localized to that country. We classified a domain as localized using \texttt{.au} for Australia, \texttt{.uk} or \texttt{.scot} for the United Kingdom, \texttt{.ca} for Canada, \texttt{.no} for Norway, \texttt{.sg} for Singapore, \texttt{.my} for Malaysia, \texttt{.jp} for Japan, \texttt{.kr} for South Korea, \texttt{.ng} for Nigeria, \texttt{.cn} for China, and \texttt{.in} for India. The United States has no equivalent country-level domain, so we instead classified a source as localized if it used a \texttt{.gov} or \texttt{.edu} domain and had no country-level domain, or belonged to a set of major American health institutions (the Cleveland Clinic, Johns Hopkins, the Mayo Clinic, and the CDC).

To identify globalized, institutional sources, we separately flagged domains belonging to major international health and humanitarian organizations, including the World Health Organization, the Pan American Health Organization, and UNICEF.

\subsection*{Regression Analysis of Google's Source Localization}

To test whether country and query topic predict how localized a search result's citations are, we regressed the share of localized citations on country and topic using ordinary least squares (OLS) regression. For each query, we computed the percentage of cited domains that were either localized to the searching country or belonged to a major United States or global institutional source (a \texttt{.gov} or \texttt{.edu} domain, one of the named American health institutions, or one of the international health organizations described above), and used this percentage as the dependent variable. Queries with no citations were excluded, and the analysis was restricted to English-language queries.

We modeled this outcome as a function of country and topic, with the United States and cancer as the respective reference levels, so that all reported coefficients are interpreted as differences from these baselines. Because topic and query section (the broader health category, such as common conditions or Traditional Chinese Medicine) are collinear, including both in the same model leaves the design matrix rank deficient, and section carries no information beyond what the topic dummies already encode. We therefore fitted two separate specifications, a primary model with topic and country as predictors, and a secondary model that replaces topic with section. We report standard errors using the HC3 estimator in both specifications.

\subsection*{Zero-shot Labeling}

For each AI Overview response returned by Baidu (China) and Google (United States), we applied a zero-shot probabilistic labeling procedure using a locally deployed Llama-3.1-70B model run through Ollama. Large pretrained language models show strong zero-shot performance on classification tasks framed as natural language prompts, drawing on instruction following and prompt-based reasoning to assign accurate labels from a task description alone \cite{brown2020language, sanh2022multitask}.

The labeling prompt (Supplement B) asked the model to return a structured JSON object scoring each response on five dimensions: the probability of a supportive stance toward Traditional Chinese Medicine, the probability of a skeptical stance toward Traditional Chinese Medicine, the probability that the response recommended consulting a healthcare professional, the probability that it recommended an emergency referral (for example, calling emergency services or visiting an emergency room), and whether it included an explicit safety warning. All labeling was conducted in English. Chinese-language responses were analyzed for their semantic content regardless of source language.

Zero-shot labels were manually reviewed by a reviewer fluent in both English and Mandarin Chinese. For each of the five features, 10 observations with low values and 10 with high values were randomly sampled for manual review. Agreement between the zero-shot labels and manual review is reported in Supplement C.

\section*{Acknowledgments}
We acknowledge the Stamps Foundation and Dartmouth College for their support of this research.

\section*{Competing Interest}
The authors declare that they have no competing interests.

\section*{Funding}
The authors received no external funding for this work.

\section*{Author Contributions}
Mingyue Zha: conceptualization, investigation, formal analysis, visualization, manuscript writing. Herbert Chang: conceptualization, supervision, and manuscript editing.

\section*{Data Availability}
The data underlying this article, including query logs, citation extractions, and zero-shot labeling outputs, will be made available upon publication at a repository to be determined, and can be accessed with a unique identifier to be assigned at that time.

\section*{Ethics Statement}
This study analyzed publicly accessible outputs of commercial AI Overview systems and did not involve human participants or personally identifiable information.

\bibliographystyle{unsrtnat}
\bibliography{references}

\end{document}